\documentclass[showpacs,twocolumn,prb,floatfix,amssymb,aps,superscriptaddress]{revtex4}
\usepackage{graphicx}

\begin{document}
\hsize\textwidth\columnwidth\hsize\csname@twocolumnfalse\endcsname

\title{Low-energy properties of anisotropic two-dimensional spin-1/2
Heisenberg models in staggered magnetic fields}

\author{Bin Xi}
\affiliation{Department of Physics, Renmin University of China, Beijing
100872, China}
\affiliation{Institute of Theoretical Physics, Chinese Academy of Sciences,
Beijing 100190, China}
\affiliation{College of Physical Sciences, Graduate University of the
Chinese Academy of Sciences, Beijing 100049, China}

\author{Shijie Hu}
\affiliation{Institut f\"ur Theoretische Physik, Georg-August-Universit\"at
Goettingen, 37077 G\"ottingen, Germany}
\affiliation{Department of Physics, Renmin University of China, Beijing
100872, China}

\author{Jize Zhao}
\affiliation{Institute for Solid State Physics, University of Tokyo, Kashiwa,
Chiba 277-8581, Japan}
\affiliation{Department of Physics and Astronomy, California State University,
Northridge, CA 91330, USA}

\author{Gang Su}
\affiliation{College of Physical Sciences, Graduate University of the
Chinese Academy of Sciences, Beijing 100049, China}

\author{B. Normand}
\affiliation{Department of Physics, Renmin University of China, Beijing
100872, China}

\author{Xiaoqun Wang}
\affiliation{Department of Physics, Renmin University of China, Beijing
100872, China}

\date{\today}

\begin{abstract}

We present a systematic study of the anisotropic spin-1/2 Heisenberg model
in staggered magnetic fields in two dimensions (2D). To mimic real materials,
we consider a system of coupled, antiferromagnetic chains, whose interchain 
interaction can be either ferro- or antiferromagnetic. When the staggered 
field is commensurate with the magnetic interactions, an energy gap opens 
immediately and follows a power law as a function of the applied field, 
similar to the situation in 1D. When the field competes with the 
interactions, a quantum phase transition (QPT) occurs from a gapless, 
magnetically ordered phase at low fields to a gapped, disordered regime. 
We use a continuous-time Monte Carlo method to compute the staggered moment 
of the ordered phases and the spin gap of the disordered phases. We deduce 
the phase diagrams as functions of the anisotropy ratio and the applied field, 
and calculate the scaling behavior of the models in both quantities. We show 
that in the competitive case, the staggered field acts to maintain a regime 
of quasi-1D behavior around the QPT, and we discuss as a consequence the 
nature of the crossover from 1D to 2D physics.

\end{abstract}


\maketitle

\section{Introduction}

The spin-1/2 antiferromagnetic Heisenberg model in a staggered magnetic
field has attracted increasing interest as the number of real materials
it describes continues to rise. While the model in zero field is gapless
in all dimensions, the staggered field lowers symmetries, changes
universality classes, and induces gapped states and unconventional,
gapped elementary excitations. Intensive investigations into staggered-field
effects in quantum magnets were motivated originally by inelastic neutron
scattering experiments on the quasi-1D spin chain copper
benzoate.\cite{Dender1} Application of a uniform magnetic field to this
material caused an unexpected gap to open in the excitation spectrum.
This energy gap follows a power law as a function of the applied magnetic
field, but its magnitude depends strongly on the direction of the applied
magnetic field.

This unusual experimental finding was soon explained\cite{Oshikawa1,Affleck1}
by the fact that copper benzoate is not a perfect chain, but has an
alternating crystal structure giving rise to a staggered gyromagnetic tensor,
possibly accompanied by a Dzyaloshiskii-Moriya (DM) interaction. The full
Hamiltonian in a uniform external magnetic field is
\begin{eqnarray}
\hat{H} = \sum_i & & \!\!\! \left[ J \hat{\bf S}_i\cdot \hat{{\bf S}}_{i+1}
 - (-1)^i {\bf D} \cdot \hat{\bf S}_i \times \hat{\bf S}_{i+1} \right.
\nonumber \\ & & \left. - \mu_B {\bf H} \cdot \left( {\bf g}^u + (-1)^i
{\bf g}^s \right) \cdot \hat{\bf S}_i \right],
\label{HSDMZ}
\end{eqnarray}
where the three terms are respectively the antiferromagnetic Heisenberg
interaction, the DM interaction, and the Zeeman splitting energy. In
addition to the superexchange coupling $J > 0$, ${\bf D}$ is the DM
vector, ${\bf H}$ the external field, and ${\bf g}^{u,s}$ are the
uniform and staggered components of the gyromagnetic tensor. By making
a local gauge transformation, which rotates the spins on two separate
sublattices, and by neglecting all contributions at higher orders in
$D/J$, Eq.~(\ref{HSDMZ}) can be mapped to the simplified Hamiltonian
\begin{eqnarray}
\hat{H} = \sum_i & & \!\!\!\! \left[ J \hat{\bf S}_i \cdot \hat{{\bf S}}_{i+1}
 - H S_{i}^{x} - (-1)^{i} h_s S_{i}^{z} \right],
\label{HSDMS}
\end{eqnarray}
where $h_s$ is an effective staggered field proportional to the product of 
$H$ with a linear combination of $D$ and $g^s$; novel features therefore 
arise only when such a material is subject to an external magnetic field. 
The Hamiltonian of Eq.~(2) can be mapped (by neglecting Zeeman splitting in
the $z$-direction) into a sine-Gordon model, a minimal framework whose
bosonized version provides a good description of the opening of the spin
gap.\cite{Oshikawa1,Affleck1,Oshikawa2,Essler1,Essler2,Essler3,Kuzmenko1}
Further unconventional features of this model include the specific
heat\cite{Essler2}, magnetic susceptibility,\cite{Affleck1,Wolter2}
dynamical structure factor,\cite{Essler1,Essler3,Kuzmenko1} line shape
in electron spin resonance (ESR) measurements,\cite{Oshikawa2}
magnetization,\cite{Wolter1,Wolter2} and the presence of mid-gap
states.\cite{Lou1,Zhao1,Lou2,Lou3}

While these numerous studies assumed that the second term in Eq.~(2) has
no significant role for weak magnetic fields, in fact the uniform and
staggered components may compete. A complete description of systems with
DM interactions in arbitrary fields still requires the full exploration of
the original Hamiltonian (1). A systematic investigation by Wang and
coworkers\cite{Zhao1} using the density-matrix renormalization-group
(DMRG) technique showed that the low-energy, high-field properties of
copper benzoate are dominated by the uniform magnetic field, on which
the spin gap depends linearly, while a crossover regime exists at
intermediate fields. This prediction was later confirmed by ESR
experiments in applied fields up to 35 T.\cite{Nojiri1}

Most staggered-field studies have focused on materials which are almost
ideally 1D in nature, such as copper benzoate,\cite{Dender1,Asano1,Ajiro1,
Asano2,Nojiri1} copper pyrimidine,\cite{Feyerherm1,Wolter1,Wolter2,Zvyagin1,
Zvyagin2} $\text{Yb}_4\text{As}_3$,\cite{Kohgi1,Oshikawa3,Shiba1} and
$\text{KCuGaF}_6$.\cite{Izumi1} However, real materials always have some
interchain coupling, which in some cases may be comparable to the staggered
magnetic field, if not also to the intrachain coupling. How the interchain
interaction may change the essential physical properties remains an open
question. Recent experiments on the weakly coupled chain system
$\text{CuCl}_2 \cdot$2(dimethylsulfoxide) (CDC)\cite{Kenzelmann1,chen1}
indicate that a gap opens at a finite value of $h_s$, rather than at zero,
when the uniform magnetic field is applied; the power-law dependence of
the excitation gap seems to be different from that observed in quasi-1D
materials. Early attempts to understand this behavior include calculations
for the two-leg ladder,\cite{wang1,wang2} which show a QPT taking place as
a consequence of the competition between the staggered magnetic field and
the interchain coupling. A chain mean-field theory developed\cite{Sato1}
to study the spin gap as function of the staggered field in 2D and 3D found
a spin gap opening immediately with the applied field (the ``noncompetitive
case'' defined in Sec.~II and discussed in Sec.~IV), a conclusion confirmed
by DMRG.\cite{Zhao2}

In a system where the staggered field competes with the ordering pattern
favored by the magnetic interactions, a finite value of $h_s$ is required
to induce a QPT. To date the only results available for this case are at
the mean-field level, and may not deliver reliable conclusions for 2D
systems. In particular, the linear dependence of the excitation gaps on the
magnetic field appears to be inconsistent with experiment.\cite{Kenzelmann1}
In this paper we contribute to the understanding of coupled Heisenberg chains 
in staggered magnetic fields by performing continuous-time Monte Carlo 
simulations using the worm algorithm,\cite{prokof1,prokof2,prokof3} and 
use our results to discuss the influence of interchain interactions and 
staggered fields on the ground state and the low-energy excitations.

This paper is organized as follows. In Sec.~II we provide a formal
introduction to the model and to the Monte Carlo method. In Sec.~III we
test our numerical techniques by computing the staggered magnetization
in zero field, making contact with known results for coupled-chain systems 
and the square lattice. In Sec.~IV we present our Monte Carlo results for 
the noncompetitive case, which demonstrate the 2D nature of the system. 
Section V contains our complete results for the competitive case, including 
the determination of the QPT, the staggered magnetization in the ordered 
phase, the gap in the disordered phase with corresponding fitting exponents, 
and a comparison of the phase diagram with different mean-field theories.
We summarize our investigation in Sec.~VI.

\section{Model and Method}

In this study we investigate the $S = 1/2$ Heisenberg model on a spatially
anisotropic square lattice in a staggered magnetic field. We stress that the
purpose of our analysis is to determine, both qualitatively and quantitatively,
the effects of a staggered field in different geometries. We do not aim to
make a direct comparison with experiment, but rather to demonstrate the
physical properties which may motivate the search for and characterization
of an appropriate material in this class. Thus we focus here on effective
models without the uniform magnetic field. For completeness we consider
both the ``noncompetitive'' case, where the geometry of the magnetic
interactions and the staggered field are commensurate, and the ``competitive''
case, where they are not. We consider an antiferromagnetic intrachain coupling
with both ferromagnetic and antiferromagnetic interchain couplings, which we
will label respectively as $(\pi,0)$ and $(\pi,\pi)$. The two different
spatial arrangements of the staggered field, also $(\pi,0)$ and $(\pi,\pi)$,
then give one competitive and one noncompetitive situation for each case.

\begin{figure}[t]
\includegraphics[width= 7.5cm, clip]{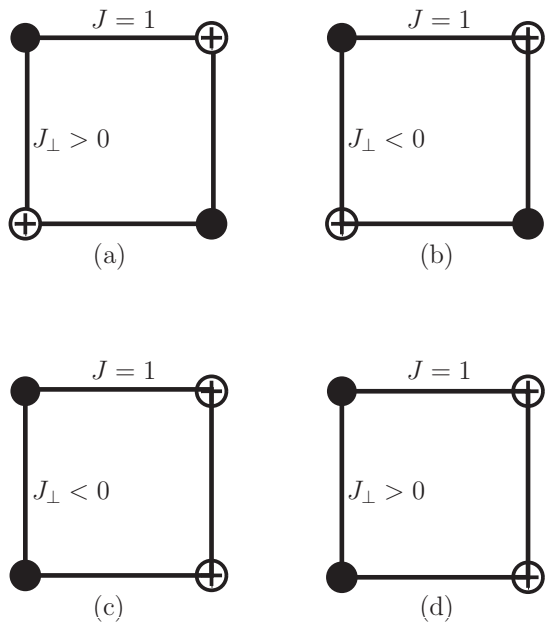}
\caption{Representation of couplings and staggered fields for the four
possible cases considered here. Dots denote a field positive along $z$
and crosses a negative field. The magnetic interactions have $(\pi,\pi)$
geometry in cases (a) and (d), and $(\pi,0)$ geometry in cases (b) and
(c). The applied field geometry is $(\pi,\pi)$ in cases (a) and (b),
which are described by Eq.~(\ref{H1}), and $(\pi,0)$ in cases (c) and (d) 
[Eq.~(\ref{H2})]. Cases (a) and (c) are therefore non-competitive, while 
(b) and (d) are competitive.}
\label{model}
\end{figure}

The Hamiltonians can be expressed in the forms
\begin{eqnarray}
\hat{H}_{1} & = & \sum\limits_{i,j=1}^L J \hat{S}_{i,j} \cdot \hat{S}_{i+1,j}
 + \sum\limits_{i=1}^L J_\bot \hat{S}_{i,j} \cdot \hat{S}_{i,j+1} \nonumber\\
 & & + \sum\limits_{i=1}^L (-1)^{i+j} h_s S_{i,j}^z, \label{H1} \\
\hat{H}_{2} & = & \sum\limits_{i=1}^L J \hat{S}_{i,j} \cdot \hat{S}_{i+1,j}
 + \sum\limits_{i=1}^L J_\bot \hat{S}_{i,j} \cdot \hat{S}_{i,j+1} \nonumber\\
 & & + \sum\limits_{i=1}^L (-1)^i h_s S_{i,j}^z, \label{H2}
\end{eqnarray}
where $L$ is the linear dimension of the lattice and $J$ is the intrachain
coupling, which we take as the unit of energy. The interchain coupling is
$J_\bot$, $h_s$ is the magnitude of the effective staggered magnetic
field, and $\hat{S}_{i,j}$ is the spin operator at lattice site $(i,j)$,
with $i$ the index in the chain direction ($x$) and $j$ the interchain
index ($y$-direction). The four situations are represented schematically
in Fig.~\ref{model}, where the sign of $J_\bot$ determines the nature of
the interchain interaction and the dots or crosses correspond to up and 
down orientations of the applied staggered field. Figures \ref{model}(a) 
and (b) represent respectively the non-competitive and competitive cases
of model (\ref{H1}), while Figs.~\ref{model}(c) and (d) represent the
non-competitive and competitive cases of model (\ref{H2}).

For our investigation of the ground states and lowest excitations of these
models, we use the continuous-time worldline quantum Monte Carlo (QMC) method
with ``worm'' updates. The concept of continuous-time worldlines was first
induced in QMC by N.~V.~Prokof'ev and coworkers in 1996.\cite{prokof1} 
Unlike the standard, discrete-time QMC algorithms, which are based on the 
Suzuki-Trotter decomposition, in the continuous-time method the Hamiltonian 
$\hat{H} = \hat{H}_0 + \hat{V}$ is separated into a diagonal term and a 
perturbation term. For a time-dependent perturbation, the partition function 
$Z = \text{Tr} ( e^{-\beta \hat{H}})$ can be expressed as $Z = \text{Tr} 
[e^{-\beta \hat{H}_0} U(\beta)]$, with $U(\beta)$ the Matsubara evolution 
operator, 
\begin{eqnarray}
U(\beta) & = & 1 - \int_0^\beta d\tau_1 \hat{V}(\tau_1) U(\tau_1) \nonumber \\
 & = & 1 - \int_0^\beta d\tau_1 \hat{V}(\tau_1) \\
 & & + \int_0^\beta d\tau_1 \int_0^{\tau_1} d\tau_2 \hat{V}(\tau_1) 
\hat{V}(\tau_2) U(\tau_2) \nonumber \\
 & = & (-1)^n\sum_{n=0}^\infty \int_0^\beta \!\! d\tau_1 \dots \!\!
\int_0^{\tau_{n-1}} \!\!\!\!d\tau_{n-1} \hat{V}(\tau_1) \dots 
\hat{V}(\tau_n), \nonumber
\end{eqnarray}
in which $V(\tau) = e^{\tau H_0} V e^{-\tau H_0}$. In this series, each
integral corresponds to a worldline configuration with $n$ ``kinks,''
located at the points $0 < \tau_n < \tau_{n-1} < \dots < \tau_1 < \beta$
and varying continuously in the imaginary-time direction.\cite{prokof1}
The integrals are evaluated by Monte Carlo sampling of the kink configurations.
In contrast to discrete-time QMC algorithms, there is no systematic error
caused by imaginary-time discretization. 

Based on this continuous-time worldline formulation, Prokof'ev and coauthors 
also developed an update algorithm based on breaking a worldline by inserting 
a pair of creation and annihilation operators, which then evolve by local 
moves.\cite{prokof2} (This became known as the ``worm'' update from the 
motion of pairs of worldline kinks.) The algorithm considers two configuration 
spaces, denoted Z and G: while Z contains only closed worldlines, and is the 
space of the partition function, G contains a worldline broken by a physical 
process connecting points $(\vec{r}_{i}, \tau_i)$ and $(\vec{r}_{j},\tau_j)$. 
The Z and G configuration spaces can be exchanged by creation or annihilation 
of an $(i,j)$ ``worm'' pair on the same flat worldline. This is equivalent to 
processes in the grand canonical ensemble. In the G configuration space, 
processes which move the worms in both time (changing worm position in the 
imaginary time direction on the same site) and space (the worm is moved to 
a neighboring site, creating or annihilating a kink) update the configuration 
and change its structure. 

In the Z configuration space it is easy to compute thermodynamic quantities 
such as energy, magnetization, susceptibility, and specific heat. In the G 
configuration space, each accepted worm move results in a contribution to 
the histogram of the Green function, $G(\vec{r}_{i} - \vec{r}_{j},\tau_{i}
 - \tau_{j})$. With sufficiently many Monte Carlo steps, one obtains a 
convergent Matsubara Green function, defined as
\begin{equation}
G(r,\tau) = G(\vec{r}_{i} - \vec{r}_{j},\tau_{i} - \tau_{j}) = \langle 
\mathcal{T}_\tau S^+_{\vec{r}_{i}}(\tau_{i}) S^-_{\vec{r}_{j}}(\tau_{j}) 
\rangle,
\end{equation}
where $\mathcal{T}_\tau$ is is the time-ordering operator. The average 
$\langle \dots \rangle$ is obtained from the histogram, and is normalized 
by $G(0,0)$. From the statistical average of the Green function, one may 
obtain the critical exponents of $G$, the single-particle excitation 
spectrum, and certain correlation functions.

In comparison with previous results by the DMRG technique,\cite{Zhao2}
QMC algorithms have the major advantage of treating 2D models. In the
continuous-time worldline approach, lattice sizes of $L$ up to 100 and
inverse temperatures $\beta = 1/T$ up to 100 can be accommodated in 
approximately 24 hours of calculation time when working on a server 
with an Inter XEON E5460 CPU. A further advantage of QMC is that it is
easy to work with periodic boundary conditions, so that there are no
difficulties caused by edge states.\cite{Zhao2} We therefore expect
that reliable numerical results can be obtained for the 2D case by
this technique. In this paper, we focus on the energy gap $\Delta$,
which characterizes gapped phases, and the transverse staggered 
magnetization $M_\bot^s$, which is characteristic of ordered phases. 
The gap is obtained\cite{prokof3} from the Green function through the 
expression
\begin{equation}
\Delta = -\frac{\ln[G(p,\tau)/G(p,\tau_0)]}{\tau - \tau_0},
\label{ed}
\end{equation}
while the staggered correlation function 
\begin{equation}
C_{\bf Q}^s (r) = \langle \hat{S}^+_0 \hat{S}^-_{\bf r} e^{i {\bf Q.r}} 
\rangle 
\label{escf}
\end{equation}
is obtained from the Green-function histogram. The staggered magnetization 
is defined as $M_\bot^s = \sqrt{\sum_{\bf r} C_{\bf Q}^s (r)}/N$, where 
$\textbf{Q} = (\pi,\pi)$ for model (\ref{H1}), and $\textbf{Q} = (\pi,0)$
for model (\ref{H2}). 

\section{Magnetization at zero field}

\begin{figure}[t]
\includegraphics[width=8.5cm, clip]{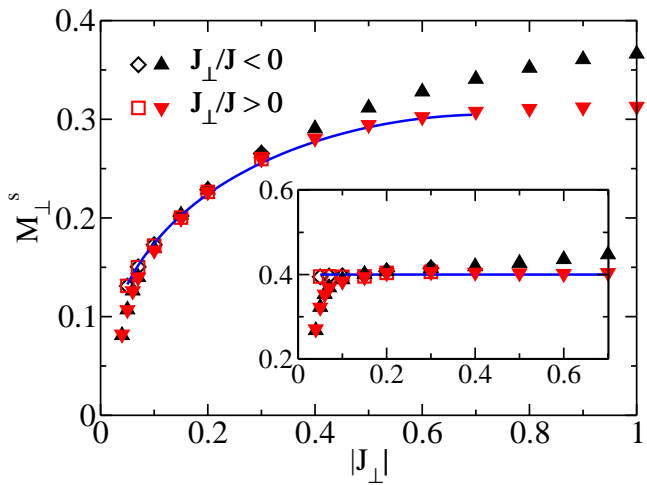}
\caption{(Color online) Staggered magnetization $M_\bot^s$ as a function of
interchain coupling for both signs of $J_\bot$, at zero staggered field.
The isotropic square lattice is represented by $J_\bot = 1$. Inset:
$M_\bot^s$ as a function of $- \sqrt{|J_\bot|} (1 + 0.095 |J_\bot|) \ln^{1/3} 
|J_\bot/1.3|$,\cite{ras} showing the strong logarithmic corrections to the 
square-root form of a chain mean-field theory. Up- and down-pointing triangles 
denote data obtained for $L$$\times$$L$ samples with $L = 100$, while squares 
and diamonds denote data obtained on rectangular samples with aspect ratio 
$L_x/L_y = 8$. The solid line is the fitting function deduced in 
Ref.~[\onlinecite{ras}].}
\label{hs0}
\end{figure}

We begin by investigating the case $h_s = 0$, which also serves to benchmark
our method for accuracy and systematic errors. We consider both the spatially
anisotropic antiferromagnetic Heisenberg square lattice and the lattice of
Heisenberg chains with ferromagnetic interchain coupling. For $|J_\bot|/J > 
0.15$, the lattice size $L$ in our calculation is $100$, with $\beta$ set 
equal to $L$. For lower values of the interchain coupling, we find that 
rectangular lattices are required to achieve reliable results, and we adjust 
the cluster aspect ratio accordingly. The staggered magnetization we 
compute for the 2D antiferromagnetic Heisenberg model is shown in
Fig.~\ref{hs0}. For benchmarking purposes, in the isotropic square lattice
($J = J_\bot = 1$) we find $M_\bot^s = 0.313 \pm 0.006$, which agrees very 
well with the known value of 0.306.\cite{David} The case of ferromagnetic 
interchain coupling is not entirely symmetrical, the non-universal behavior 
at larger values of $|J_\bot|$ being manifest as a more rapid saturation in 
the antiferromagnetic case. 

The problem of the weakly coupled Heisenberg spin chains is a fundamental 
one in quantum magnetism, and has received a great deal of attention over 
the last five decades. It encapsulates the physics of the crossover from 
truly 1D systems, dominated by quantum fluctuation effects, to 
high-dimensional, renormalized classical behavior. Because the $S = 1/2$ 
Heisenberg chain is already critical, with a gapless ground state, any 
transverse coupling in an unfrustrated geometry is thought to give rise 
to magnetic order, and extensive investigation has reinforced the general 
agreement that the critical $J_{\bot c} = 0$. Early discussion of the 
critical behavior of the ordered moment and N\'eel temperature used the 
high-dimensional framework of renormalized spin waves, and suggested a
purely logarithmic rise of $M_\bot$ with $|J_\bot|$.\cite{ro} By contrast, 
in a chain-based weak-coupling approach,\cite{rs} where the transverse 
interactions are modeled as an effective staggered magnetic field, the 
ordered moment has a power-law dependence, $M_\bot \propto \sqrt{|J_\bot|}$, 
albeit with suspected logarithmic corrections. The definitive study of the
problem was performed by Sandvik,\cite{ras} using QMC within a multichain 
mean-field theory, and reveals strong logarithmic corrections to the 
square-root dependence, $M_\bot \propto - \sqrt{|J_\bot|} \ln^{1/3} 
|J_\bot|$ (with a weak additive linear term). The power of 1/3 in the 
logarithm was also found in the problem of the single chain in a 
staggered field.\cite{Affleck1} The importance of the logarithmic terms 
lies in the presence of marginally irrelevant operators, which are neglected 
in the transformation of the high-dimensional Heisenberg model to a chain 
in an effective staggered field. 

Our data for both signs of the transverse coupling, shown in Fig.~\ref{hs0}, 
have the same form at small $|J_\bot|$. Our results are fully consistent 
with those of Ref.~[\onlinecite{ras}]. First, we confirm our sensitivity 
to the expected logarithmic corrections, which will be essential in the 
sections to follow. Second, our results for square lattices deviate from 
the expected form at values of the transverse coupling below $|J_\bot| 
\approx 0.15$. We simulate instead rectangular systems of different aspect 
ratios up to $L_x / L_y = 8$,\cite{ras} which allows us to obtain accurate 
values of $M_\bot$ at least to $|J_\bot|/J = 0.05$, below which the 
calculations become very time-consuming. This study of the staggered 
magnetization at $h_s = 0$ therefore allows us to benchmark the accuracy 
of our results for highly anisotropic systems. The effect of the finite 
temperature in our calculations is not thought to be significant.\cite{ras} 
Quantifying the logarithmic corrections to the chain mean-field picture in 
this way will be important in Sec.~V, where we will use it as a 
guide to understanding our numerical results in the presence of a staggered 
field. We will show that the field acts to alter significantly the effective 
dimensionality of the system and the relevance of these logarithmic terms.

\section{Noncompetitive case}

In the two noncompetitive cases, the magnetic interactions and the applied
staggered field have the same spatial pattern [Figs.~1(a) and (c)]. For 
any finite value of $J_\bot$, the system is magnetically ordered at
$h_s = 0$, by a spontaneous breaking of the SU(2) spin symmetry, and has
massless spin-wave excitations. When a commensurate staggered field is 
applied, the symmetry is broken explicitly as the field direction selects 
the spin orientation. As for a ferromagnet in a uniform field, the staggered 
field serves as an anisotropy term, which opens a spin-wave gap for any finite 
value of $h_s$. In the conventional mean-field approach,\cite{Sato1} a 
standard linear spin-wave approximation gives a gap
\begin{equation}
\Delta \approx \sqrt{4 S J h_s [1 + (h_s / 4 S J)]}.
\label{encge}
\end{equation}
By extrapolation from smooth and rapidly convergent DMRG results for
antiferromagnetic Heisenberg ladders in staggered fields, the authors
of Ref.~[\onlinecite{Zhao2}] also obtained a 2D field-dependence of $\Delta
 = (2.27 \pm 0.01)\ h_s^{0.50 \pm 0.01}$. However, the noncompetitive geometry 
may also be discussed within a chain mean-field theory, where the interchain 
interactions are treated as an effective staggered field, $h = - 2 J_\bot 
M_\bot (h)$, \cite{rs} which is reinforced by the applied staggered field. 
In this treatment, the dominant physics is the breaking of the continuous 
symmetry of the spin direction, and the exact dependence on $J_\bot$ is not 
clear; that the gap $\Delta$ is independent of $J_\bot$ in Eq.~(\ref{encge}) 
underlines the fact that it is in essence purely an effect of the field on 
the chain, which would suggest some influence of 1D physics.

\begin{figure}[t]
\includegraphics[width= 8cm, clip]{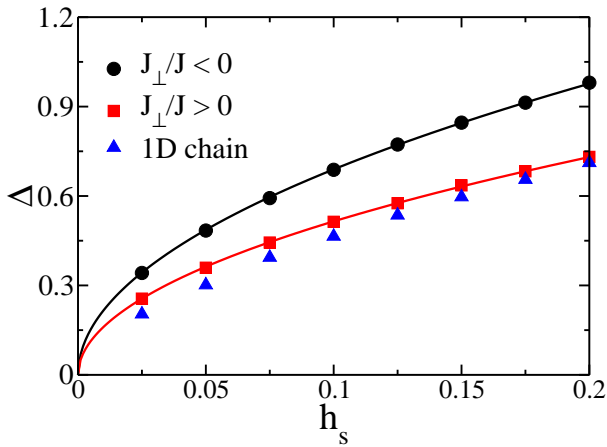}
\caption{(Color online) Excitation gap as a function of staggered
magnetic field for the 2D spin-$\frac{1}{2}$ Heisenberg square lattice
($J_\bot = J$) in the noncompetitive case. Monte Carlo results are
presented as filled circles for the $(\pi,\pi)$ geometry [model (\ref{H1})]
and as squares for $(\pi,0)$ [model (\ref{H2})]. The solid lines are fitting
curves. Shown also (triangles) are Monte Carlo results for the Heisenberg
chain.}
\label{non-competition-1}
\end{figure}

Our analysis is the first direct numerical calculation of the gap evolution
in this case. We begin by considering the ``isotropic'' square lattice 
($|J_\bot| = J$) with systems of $L = 100$ and temperatures $\beta = 100$, 
to maximize the reliability of our data by systematic extrapolation. In 
Fig.~\ref{non-competition-1} we show the dependence of the gap on the 
staggered field for both noncompetitive cases. A fit to the formula
\begin{equation}
\Delta = a_0 h_s^{\alpha}
\label{GAPSCALING}
\end{equation}
yields excellent results (Fig.~\ref{non-competition-1}), with $\Delta = 2.19
h_{s}^{0.50 \pm 0.01}$ for the $(\pi,\pi)$ case [Fig.~1(a)] and $\Delta =
1.65 h_{s}^{0.50 \pm 0.01}$ for the $(\pi,0)$ case [Fig.~1(c)]. The
exponents in both cases agree perfectly with the result predicted both by
mean-field theory and by extraolation from Heisenberg ladders. The isotropic 
square lattice shows unambiguously 2D behavior, with an immediate opening 
and square-root growth of the gap.

\begin{table}[b]
\caption{Fitting coefficients and exponents for five data sets with
different point spacings $\delta h_s$.}
\begin{ruledtabular}
\begin{tabular}{c|c|c}
\multicolumn{3}{c}{$J_\bot/J = 0.1$} \\ \hline
  $\delta h_s$ & $a_0$  & $\alpha$  \\ \hline
0.005 & 1.351 & 0.535\\
0.01  & 1.415 & 0.548\\
0.05  & 1.512 & 0.574\\
0.1   & 1.522 & 0.583\\
0.2   & 1.554 & 0.624
\end{tabular}
\end{ruledtabular}
\label{T1}
\end{table}

\begin{figure}[t]
\includegraphics[width= 8cm, clip]{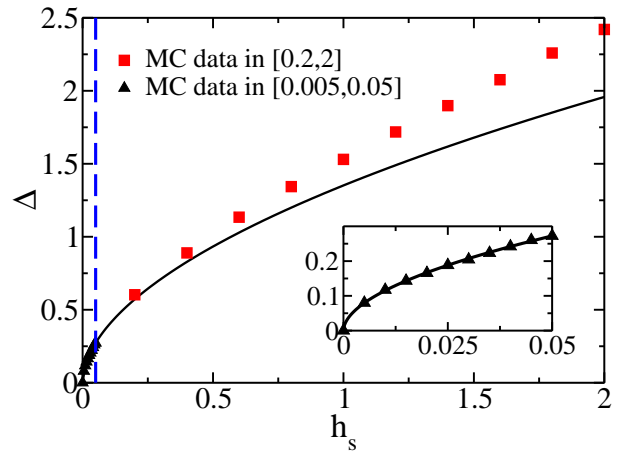}
\null
\caption{(Color online) Excitation gap as a function of staggered field
for the anisotropic 2D spin-$\frac{1}{2}$ antiferromagnetic Heisenberg model
in the noncompetitive case for $J_\bot/J = 0.1$. Triangles and squares 
correspond respectively to $h_s$ steps of 0.005 and 0.2. The fitting curve 
is obtained from the data with $\delta h_s = 0.005$ (inset). The blue, 
dashed line marks the 2D regime for $J_\bot/J = 0.1$. }
\label{test}
\end{figure}

By contrast, in the purely 1D case is it known\cite{Oshikawa1,Affleck1}
that the gap opens according to $\Delta \propto h_s^{2/3}$ with 
logarithmic corrections. This behavior is shown as the triangles in 
Fig.~\ref{non-competition-1}. The more rapid growth of the gap in 2D may 
be considered heuristically as the consequence of a mutual reinforcement 
of the applied and effective staggered fields, the latter arising from the 
magnetic moment enhanced by the former. The key question to address is 
whether the system displays any kind of continuous crossover, as a 
function of $|J_\bot|$, from a regime characterized by 1D exponents to a 
2D regime. By taking the non-competitive $(\pi,\pi)$ case at  $J_\bot =
0.1$ and fitting the gap obtained from ten sets of data points with five
different $h_s$ intervals, 0.005, 0.01, 0.05, 0.1, and 0.2, we find the
fitting parameters listed in Table \ref{T1}. It is clear that the
exponent changes from a 2D form to a 1D form: for large $h_s$, the system
is effectively no longer aware of the coupling, and has 1D behavior, while
at sufficiently small $h_s$ the 2D behavior always emerges. The same result
is illustrated in Fig.~\ref{test} using the data points with $\delta h_s =
0.005$ and $\delta h_s = 0.2$. It is clear that the latter data set does
not fall on the square-root curve obtained from the former. We may conclude
that the 2D regime in this model is given approximately by $h_s < J_\bot/2$.

\section{Competitive case}

We turn now to the competitive case, where the interchain coupling and the
staggered field compete to establish the pattern of magnetic order. From a
mean-field analysis of this system,\cite{Sato1} there exist two different
phases at finite $h_s$, an ordered phase with spontaneous symmetry-breaking
(SSB, of the continuous U(1) symmetry) in the plane normal to the staggered 
field and a gapped, ``symmetric'' phase in which the spins are oriented in 
the field direction. 

\begin{figure}[t]
\includegraphics[width=8.5cm]{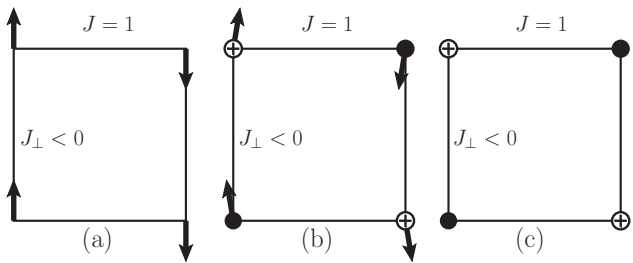}
\caption{Schematic representation of the spin state in the competitive
case on varying the staggered field $h_s$. The system has FM interchain
interactions and a $(\pi,\pi)$ staggered field [model (\ref{H1})]. Black
arrows represent the spin directions and the staggered field is represented
by dots (up) and crossed circles (down).} \label{ssb1}
\end{figure}

A phenomenological description of the situation is illustrated in 
Fig.~\ref{ssb1} for the case of a $(\pi,\pi)$ staggered field applied to 
a system with $J_\bot < 0$. At $h_s = 0$, the system is ordered with a 
spontaneous breaking of SU(2) symmetry and one-site translational 
invariance in the transverse direction. For $h_s = 0^+$, a spin-flop 
transition occurs into the plane perpendicular to the field, represented 
as the $(xy)$ plane, but there remains a SSB in this plane 
[Fig.~\ref{ssb1}(a)]. For finite $h_s$, the spins are forced to rotate
into the direction of the staggered field, or into the $yz$ plane in
Fig.~\ref{ssb1}(b), adopting a canted structure with two-site translational
invariance in the transverse direction. The finite magnetic order parameter
in this phase is reduced by the staggered field, and the excitations of the
system remain gapless. Finally, when the staggered field is increased beyond
a critical value $h_c$, the order parameter is driven to zero, long-range
order collapses, and the spins are fully oriented in the field direction
[Fig.~\ref{ssb1}(c)].

\begin{figure}[t]
\includegraphics[width=7.5cm, clip]{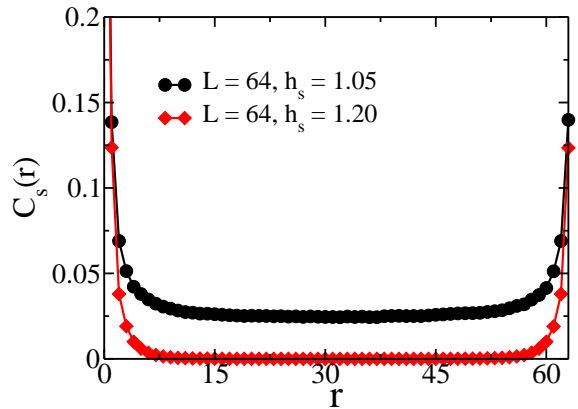}
\caption{(Color online) Staggered correlation function $C_s(r) = \langle 
\hat{S}^+_0 \hat{S}^-_{\bf r} e^{i {\bf Q.r}}\rangle$ (see text)
with $L = 64$ and $J_\bot/J = -1.0$, for two staggered fields, $h_s = 1.05\
\text{and}\ 1.2$, chosen to represent respectively the ordered and the gapped
phase.}
\label{cs}
\end{figure}

We begin by our numerical analysis by illustrating the staggered correlation 
function within the system. Figure \ref{cs} shows $C_s (r)$ as a function of 
distance for model (\ref{H1}), with $J_{\perp} = -1$ and for two values of 
$h_s$. For $h_s = 1.05$, staggered correlations are finite and there is 
long-range order in the plane transverse to the applied field. By contrast, 
for $h_s = 1.2$, $C_s (r)$ falls to zero abruptly away from the edges of the 
system; these two values of $h_s$ therefore fall on opposite sides of the 
anticipated critical staggered field, $h_c$. We use the staggered 
magnetization to analyze both the SSB phase and the field-induced quantum 
phase transition between the SSB and symmetric phases. We characterize the 
symmetric phase by its gap, defined in Eq.~(\ref{ed}), and by the scaling 
of this gap.

\subsection{Critical point} 

From the discussion above, the critical point $h_c$ is the single most
important quantity in the descriptiontion of the competitive case. Once
$h_c$ is determined, a process achieved most accurately using the behavior 
of the order parameter in the SSB phase, the gap and the 2D excitations of
the symmetric phase can be calculated with high accuracy. To determine the
critical point, or the phase boundary in the space of $J_\bot$ and $h_s$,
we apply the finite-size-scaling hypothesis to the behavior of the
staggered correlation functions $C_{\pi,0}$ and $C_{\pi,\pi}$
[respectively for models (\ref{H1}) and (\ref{H2})] as functions of $L$.
These are expected\cite{FISH1} to obey the scaling form
\begin{equation}
C(\delta) = L^{2 - d - z} f (\delta L^{1/\nu},\beta/L^z),
\end{equation}
where $\delta = h_s - h_c$, $\beta$ is the inverse temperature, and $z$ is
the dynamical exponent.

The critical point $h_c$ can be measured accurately by computing the
staggered correlation function near the critical point for different lattice
sizes $L\times L$. On fixing $\beta/L^z = 1$, the scaling function $f$
depends on a single parameter, $\delta L^{1/\nu}$. Precisely at the critical
point, $\delta = 0$, $CL^{d+z-2}$ is independent of the lattice size, which
implies that calculated curves for $CL^{d+z-2}$ as functions of $h_s$ for
different lattice sizes should cross at the critical point. It is clear
from this form that $z$ is a measure of the system coherence: as $z$
becomes larger, the coherence of the system vanishes more rapidly in
space at the critical point. Here we test that the dynamical exponent takes
the value $z = 1$, and we find this to hold at all points around the critical
region.

\begin{figure}
\includegraphics[width=8.5cm, clip]{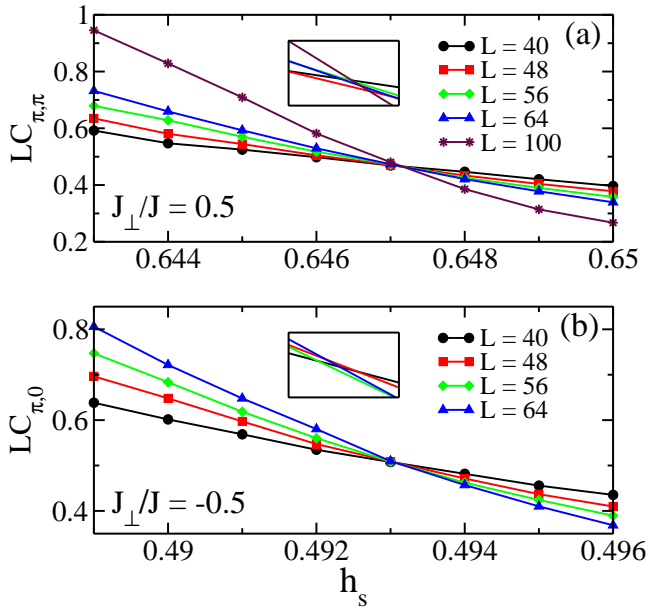}
\caption{(Color online) Determination of $h_c$ from the finite-size-scaling
hypothesis for (a) $J_\bot = 0.5$ and (b) $J_\bot = - 0.5$. Error bars 
are set by the inexact crossing of the lines, shown in the insets. In 
panel (a), $0.6471 \le h_s \le 0.6472$ and $0.462 \le LC_{\pi,\pi} \le 
0.471$. In panel (b), $0.49300 \le h_s \le 0.49315$ and $0.502 \le LC_{\pi,0} 
\le 0.510$.}
\label{cross}
\end{figure}

We have simulated 11 different values of $|J_\bot|$ for both ferro- and
antiferromagnetic interchain coupling, with fixed $L/\beta = 1$ and at least
four lattice sizes, namely $L = 40$, 48, 60, and 64. We have verified the
accuracy of our results by performing calculations with $L = 100$ for the
case $J_\bot = 0.5$. For a determination of the critical point, we work only 
with square samples at all values of $|J_\bot|$; the simultaneous vanishing 
of the superfluid density (spin stffness) in both directions at $h_c$ negates 
the advantanges of rectangular samples for low $|J_\bot|$ (Sec.~III) in this 
case. Examples of the sets of crossing curves are illustrated in 
Fig.~\ref{cross} for $J_\bot = 0.5$ and $-0.5$. The error in $h_c$ can be 
estimated from the separation of the different intersection points within 
a single manifold of curves; all errors $\delta h_c$ are of order $10^{-4}$. 
The full numerical details determined from this procedure are presented in 
Table \ref{T3}. The phase diagram deduced from these values of $h_s$ is 
discussed in Sec.~VD.

\begin{table}[b]
\caption{Critical staggered fields $h_c$ for systems with transverse couplings
$|J_\bot|$ varying from 0.05 to 1. Shown are results for both ferromagnetic
interchain couling in a $(\pi,\pi)$ staggered field (\ref{H1}) and
antiferromagnetic interchain coupling in a $(\pi,0)$ field.}
\begin{ruledtabular}
\begin{tabular}{c|c|c|c|c|c}
\multicolumn{3}{c|}{$(\pi,\pi)$} & \multicolumn{3}{c}{$(\pi,0)$}\\ \hline
  $J_\bot$ &   $h_{c}$ &$\delta h_c$ & $J_\bot$ &  $h_{c}$ &$\delta h_c$
  \\\hline
-0.05      & 0.02459  &0.00007 &0.05      & 0.02606 & 0.00008   \\
-0.1       & 0.06336  &0.00011 &0.1       & 0.06964  & 0.00027\\
-0.2       & 0.15805  &0.00015 &0.2       & 0.18495  & 0.00013\\
-0.3       & 0.26495  &0.00016 &0.3       & 0.32470  & 0.00022\\
-0.4       & 0.37775  &0.00014 &0.4       & 0.48034  & 0.00008\\
-0.5       & 0.49305  &0.00012 &0.5       & 0.64715  & 0.00015\\
-0.6       & 0.60990  &0.00006 &0.6       & 0.82166 &  0.00018\\
-0.7       & 0.72630  &0.00012 &0.7       & 1.00225  & 0.00020\\
-0.8       & 0.84300  &0.00011 &0.8       & 1.18617  & 0.00014\\
-0.9       & 0.95850  &0.00017 &0.9       & 1.37339  & 0.00015\\
-1.0       & 1.07300  &0.00021 &1.0       & 1.56290 &  0.00019
\end{tabular}
\end{ruledtabular}
\label{T3}
\end{table}

\subsection{Staggered Magnetization}

\begin{figure}
\includegraphics[width=8.5cm, clip]{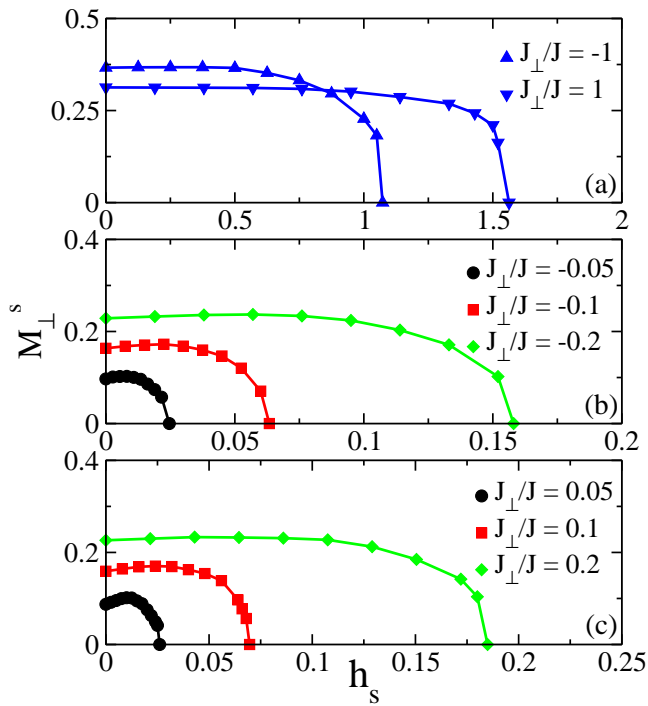}
\caption{(Color online) Transverse staggered order parameter $M_\bot^s$ as a
function of $h_s$ for different values of the interchain coupling $J_\bot$.
(a) Regime of large $|J_\bot|$, showing conventional suppression of 
$M_\bot^s$ by the competing field $h_s$. (b) and (c) Regime of small 
$|J_\bot|$, showing unconventional increase of $M_\bot^s$ with $h_s$ at 
low applied fields.}
\label{ms}
\end{figure}

For applied fields $h_s < h_c$, the SSB phase possesses a transverse
staggered moment (or off-diagonal long-range order) $M_\bot^s$. We have
calculated this quantity for the case $h_s = 0$ in Sec.~III. In general
one expects the magnetic order to be suppressed by the competing
staggered field, because of the different ordering patterns they favor,
until $M_\bot^s$ vanishes at $h_s = h_c$. In Fig.~\ref{ms} we show
$M_\bot^s$ as a function of $h_s$ for different values of the interchain
coupling $J_\bot$. For larger values of $J_\bot$, of both signs, the
order parameter shows a conventional, monotonic decrease [Fig.~\ref{ms}(a)].
However, when $|J_\bot| \lesssim 0.2$, for both signs of $J_\bot$ 
[Figs.~\ref{ms}(b) and (c)], we observe that $M_\bot^s$ first increases 
despite the increase in the competing staggered field. We stress that 
our results for small $|J_\bot|$ were obtained on rectangular samples 
with aspect ratio $L_x/L_y = 8$, following the procedure established in 
Sec.~III as providing the highest available accuracy. 

To our knowledge, this novel and purely quantum mechanical effect has not
been remarked upon previously in coupled Heisenberg chains. An analogous
effect can be found in dimerized quantum magnets\cite{rmnrs} such as
NH$_4$CuCl$_3$, where an applied uniform magnetic field, which in principle
competes with the transverse staggered order, nonetheless causes the
staggered moment to rise at the same time as the uniform polarization is
increased. A heuristic understanding of this phenomenon is that all forms
of magnetic order are in fact competing with quantum fluctuation effects
favoring disordered states. When the disordered state is suppressed by an
applied field, more ordered spin ``weight'' is available both for the
magnetization component favored by the field and for the component favored
by the magnetic interactions.

As $h_s$ is increased further, $M_\bot^s$ is suppressed for all values
of $|J_\bot|$, and it falls continuously to zero at the second-order
quantum phase transition. In this regime the finite-size effects in our 
simulations are large, even though the statistical errors are very 
small. Unlike the determination of $h_c$ itself, there is neither a 
systematic approach by extrapolation with different system sizes, which 
may allow a sufficiently accurate determination of $M_\bot^s$ close to 
$h_c$, nor a method for the accurate estimation of such errors. Thus we 
are not able to deduce the critical exponents of the transverse staggered
moment in order to characterize this side of the quantum phase transition. 
The results of Sec.~VC below suggest that some anomalous behavior may be 
expected.

\begin{figure}
\includegraphics[width=8.5cm, clip]{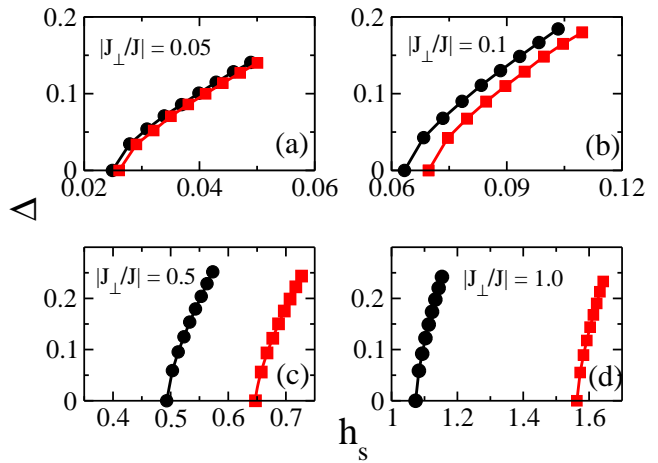}
\caption{(Color online) Gaps as a function of staggered magnetic field 
$h_s$ for four different values of $|J_\bot|$. Monte Carlo results for 
the gap are presented by black circles and red squares respectively for 
ferro- and antiferromagnetic interchain couplings. The critical points
$h_c$ are determined from finite-size scaling of the transverse order
parameter in the SSB phase. The gap is fitted to the form $\Delta = a_0 
(h_s - h_c)^\alpha$. Fitting curves are shown by solid lines and complete
fitting parameters presented in Table \ref{T2}.}
\label{competition}
\end{figure}

\subsection{Energy gap}

In the gapped, symmetric phase we wish to consider the properties of
the excited states. We compute the energy gap $\Delta$ and extract the
scaling behavior of the gap in the region $h_s - h_c < |J_\bot|/2$
deduced in Sec.~IV. In Fig.~\ref{competition} we present selected
examples of the function $\Delta (h_s)$ for a range of positive and
negative values of $J_\bot$. Unlike in the SSB phase, there are no
strong finite-size effects on the gap data close to $h_c$ for the
symmetric phase, a result we have confirmed by studies on lattices
of various sizes up to $L = 100$. For a given $h_c$, determined from
finite-size scaling of the transverse order parameter in the SSB phase,
we collect eight data points within the scaling regime $h_s - h_c <
|J_\bot|/2$. We fit our numerical data to the formula $\Delta = a_0
(h_s - h_c)^\alpha$.

\begin{table}[b]
\caption{Fitting coefficients and exponents for the excitation gap in
systems with traverse couplings $|J_\bot|$ varying from 0.05 to 1. Shown
are results for both ferromagnetic interchain coupling in a $(\pi,\pi)$
staggered field (\ref{H1}) and antiferromagnetic interchain coupling in
a $(\pi,0)$ field (\ref{H2}).}
\begin{ruledtabular}
\begin{tabular}{c|c|c|c||c|c|c|c}
\multicolumn{4}{c|} {$(\pi,\pi)$} & \multicolumn{4}{c} {$(\pi,0)$}\\ \hline
$J_\bot$ & $a_0$  & $\alpha$ &$\delta \alpha $ & $J_\bot$ & $a_0$
& $\alpha$  &$\delta \alpha$\\
\hline
-0.05     & 0.929  &  0.693 & 0.003  &0.05      & 0.943  & 0.698 & 0.004 \\
-0.1      & 0.926  &  0.718 & 0.005  &0.1       & 0.880  & 0.708 & 0.010\\
-0.2      & 0.862  &  0.724 & 0.006  &0.2       & 0.822  & 0.715 & 0.005\\
-0.3      & 0.815  &  0.718 & 0.006  &0.3       & 0.759  & 0.706 & 0.008\\
-0.4      & 0.761  &  0.704 & 0.004  &0.4       & 0.744  & 0.703 & 0.003\\
-0.5      & 0.735  &  0.700 & 0.004  &0.5       & 0.732  & 0.702 & 0.006\\
-0.6      & 0.716  &  0.693 & 0.003  &0.6       & 0.686  & 0.692 & 0.008\\
-0.7      & 0.709  &  0.694 & 0.005  &0.7       & 0.676  & 0.689 & 0.009\\
-0.8      & 0.705  &  0.692 & 0.005  &0.8       & 0.673  & 0.688 & 0.005\\
-0.9      & 0.698  &  0.692 & 0.007  &0.9       & 0.667  & 0.686 & 0.006\\
-1.0      & 0.693  &  0.692 & 0.007  &1.0       & 0.661  & 0.686 & 0.008\\
\end{tabular}
\end{ruledtabular}
\label{T2}
\end{table}

Our gap calculations are of course consistent with the values of $h_c$
computed in the SSB phase. The gaps for $J_\bot$ values of opposite
sign show only small, quantitative differences. The two gaps converge
to that of the 1D chain when $J_\bot \rightarrow 0$. As in Sec.~VA,
$h_c$ increases with the magnitude of $J_{\perp}$, reflecting the fact
that it is the competition between the staggered field and the interchain
coupling that drives the quantum phase transition. Full details of the
fitting parameters for the gap, including the errors in the exponents
we extract, are listed in Table {\ref{T2}}.

\begin{figure}
\includegraphics[width=8cm, clip]{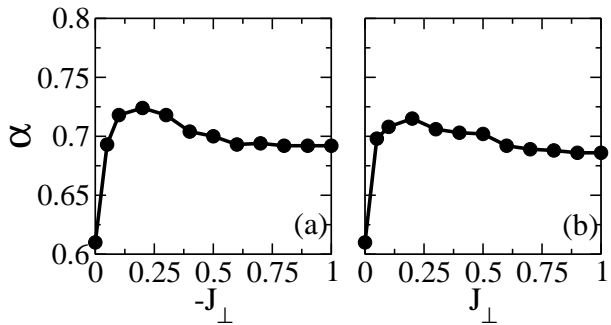}
\caption{(Color online) Critical exponents $\alpha$ of the gap $\Delta$
(Table {\ref{T2}}) as a function of $|J_{\bot}|$ for (a) ferromagnetic 
interchain coupling and (b) antiferromagnetic interchain coupling. The 
solid lines are a guide to the eye.}
\label{exp}
\end{figure}

Figure \ref{exp} shows the gap exponent extracted for different values of
$J_\bot$. Because all of the data is fitted to a simple power law, the 
logarithmic corrections in the 1D chain result in an effective exponent 
of 0.61 for $J_\bot = 0$. The exponent shows an initial increase as 
$|J_{\bot}|$ increases, but then falls weakly and remains close to 
$\alpha = 0.7$ over the remainder of the range up to $|J_{\perp}| = 1$. 
The values for the two cases of ferro- and antiferromagnetic $J_\bot$ 
differ only very slightly, and no universal value is indicated. It is 
clear, however, that the value of $\alpha$ does not fall to 0.5, even 
arbitraily close to the transition, as would be expected in a 2D mean-field 
theory or from the results of Sec.~IV. Our results therefore provide strong 
evidence for nontrivial quantum physics. 

For a heuristic explanation of these data, we appeal to the chain mean-field 
theory of Ref.~[\onlinecite{rs}], where the interchain interactions are 
treated as an effective staggered field $h_0 = - 2 J_\bot M_\bot^0$. The 
presence of a real staggered field either reinforces this effective one, 
as in the noncompetitive geometry of Sec.~IV, or suppresses it in the 
competitive geometry. Unlike the noncompetitive case, where the applied 
field breaks all continuous symmetries, in the competitive case a U(1) 
symmetry is maintained and the system may still be treated as a chain 
in a single effective field. 

At lowest order, one may write $h_{\rm eff} = - 2 J_\bot M_\bot (h_s) = 
 - 2 J_\bot M_\bot^0 + h_s$, from which $M_\bot (h_s) = M_\bot^0 (1 - h_s 
/ 2 J_\bot M_\bot^0)$. This simple linear approximation contains directly 
the competition between the applied and effective staggered fields, and 
suggests a quantum phase transition at $h_c = 2 J_\bot M_\bot^0$, where 
the applied field cancels the interactions. The quasi-linear relation 
between $h_c$ and $J_\bot$ is clear from our calculations, while the 
relevance of next-order, and possibly also logarithmic, corrections is 
clear from our results with ferro- and antiferromagnetic $J_\bot$. The 
spin chain in a positive effective staggered field is once again a problem 
with no continuous symmetry, in which a spin gap opens directly. The key 
point about this heuristic picture is that the cancellation of the applied 
and effective staggered fields results in a quasi-1D problem close to $h_c$, 
and hence the anomalous exponents we observe for all values of $J_\bot$ are 
to be expected: $h_c$ is always the staggered field required to cancel the 
2D coupling and reduce the model to a 1D system. 

The inexact values of the exponent can be ascribed to departures from 
universality in the 1D nature of this system, and are are not indicative 
of a phase genuinely intermediate between 1D and 2D. Because of the 
effectively 1D nature of the system close to the critical applied 
staggered field, logarithmic corrections to the gap scaling may remain 
significant for all values of $|J_\bot|$. We conclude these considerations 
by commenting on the coupled spin-chain system CDC,\cite{Kenzelmann1} where 
anomalous gap exponents have been measured despite the fact that the 
interchain coupling is thought to be significant. We suggest on the 
basis of our results that the origin of this behavior may lie in the 
effects of a competitive staggered field. 

\subsection{Phase diagram}

\begin{figure}
\includegraphics[width=8.5cm, clip]{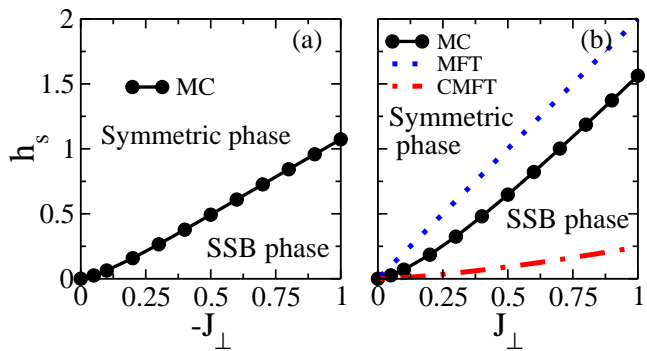}
\caption{(Color online) Phase diagrams for competitive-case models in
(a) $(\pi,\pi)$ and (b) $(\pi,0)$ staggered fields, {\it i.e.}~with
(a) ferromagnetic and (b) antiferromagnetic interchain coupling. Monte
Carlo results, given in Table \ref{T3}, are shown by solid circles and 
power-law fits (see text) by solid lines. In (b), the blue, dotted line 
marks the phase boundary predicted by a 2D mean-field theory and the red, 
dashed line that from a chain-based mean-field theory.\cite{Sato1}}
\label{phase}
\end{figure}

Finally, we present in Fig.~\ref{phase} the phase diagrams of the two
competitive-case models for ferro- and antiferromagnetic interchain
coupling. The data is taken from Table II. For comparison we show
also the chain mean-field results and 2D mean-field results of
Ref.~[\onlinecite{Sato1}]. The numerically exact boundary determined by
our quantum Monte Carlo simulations lies between those obtained from
the two mean-field theories [Fig.~\ref{phase}(b)], showing that neither
is particularly accurate for the problem of the 2D staggered magnetic
field, and confirming the general departure from universality of this
type of system.

The phase boundary, or critical point as a function of the interchain
coupling constant, can be fitted well over the full range of $J_\bot$
by a simple power law,
\begin{equation}
h_{c} = b_0|J_\perp|^\gamma,
\label{fit1}
\end{equation}
with parameters $h_{c} = 1.08 |J_\bot|^{1.15}$ for the case of ferromagnetic
interchain coupling and $h_{c} = 1.57 |J_\bot|^{1.30}$ for antiferromagnetic
coupling.

\section{Summary}

In summary, we have studied the zero-temperature phase diagram and low-energy
spin excitations of anisotropic, two-dimensional spin-$\frac{1}{2}$ Heisenberg
models on the square lattice under a staggered magnetic field. We have used
a continuous-time Monte Carlo method to calculate ordered moments and spin
gaps, and hence phase boundaries and scaling relations. At zero field, we
compute the properties of the anisotropic Heisenberg model and benchmark 
our calculations for the complexities of high anisotropy and strong 
logarithmic corrections. When the applied field and magnetic interactions 
cooperate, a gap opens immediately and scales exactly as the square root of 
the staggered magnetic field. When the two compete, we find a field-driven 
quantum phase transition from a gapless phase of staggered magnetic order, 
and characterized by spontaneous symmetry-breaking in the transverse 
direction, to a gapped, disordered, field-dominated phase. In the ordered 
phase, we find a novel enhancement of the staggered moment by a competing 
field, a purely quantum effect. We determine scaling regimes, discuss the 
scaling properties of the gap, and show that the physics of the system in 
the competitive case is essence always one-dimensional at the quantum 
critical point due to the cancellation of applied and effective staggered 
fields. 

\acknowledgments

J.~Zhao is grateful to N. Prokof'ev for sharing his continuous-time Monte
Carlo code. We acknowledge helpful discussions with G. Uhrig. J.~Zhao is 
supported in part by the Japan Society for the Promotion of Science. This 
work was supported by the National Science Foundation of China under Grant
No.~10874244 and by Chinese National Basic Research Project No.~2007CB925001.

\end{document}